\documentclass[a4paper,11pt]{article}
\usepackage{pos}
\usepackage{lineno}

\title{A search for instanton-induced decay of super-heavy dark matter in the Pierre Auger Observatory data}
\ShortTitle{Instanton-induced decay from SHDM}

\author[a]{Olivier Deligny}
\author[b]{the Pierre Auger Collaboration}

\affiliation[a]{Laboratoire de Physique des 2 Infinis Ir\`ene Joliot-Curie (IJCLab)\\
CNRS/IN2P3, Universit\'{e} Paris-Saclay, Orsay, France}
\affiliation[b]{Observatorio Pierre Auger,Av. San Martín Norte 304, 5613 Malargüe, Argentina\\
Full author list: \href{https://www.auger.org/archive/authors\_2022\_05.html}{https://www.auger.org/archive/authors\_2022\_07.html}}

\emailAdd{deligny@ijclab.in2p3.fr}
\emailAdd{spokespersons@auger.org}

\abstract{Using data collected at the Pierre Auger Observatory, we search for signatures of instanton-induced processes that would provide evidence of super-heavy particles decaying in the Galactic halo. Such particles could have been produced sufficiently during the post-inflationary epoch to match the relic abundance of dark matter inferred today. The non-observation of these signatures allows us to probe the instanton strength and to derive a bound, the best ever obtained from instanton-mediated processes, on the reduced coupling constant of gauge interactions in the dark sector: $\alpha_X \lesssim 0.09$, for $10^{9} \lesssim M_X/{\rm GeV} < 10^{19}$.}

\FullConference{%
  *** 27th European Cosmic Ray Symposium - ECRS ***\\
  *** 25-29 July 2022 ***\\
  *** Nijmegen, the Netherlands ***
}

\newcommand{\dif}{\mathrm{d}}

\def \nn {\mathbf{n}}

\begin{document}
\maketitle

Currently, the concordance model used in cosmology, the $\Lambda$CDM model where $\Lambda$ is the cosmological constant associated with dark energy and CDM stands for cold dark matter, states that the Universe is $13.8{\times}10^9$ years old and made up of $\simeq 5$\% baryonic matter, $\simeq 26$\% dark matter (DM) and $\simeq 69$\% dark energy~\cite{refId0}. To describe DM, which should be cold to allow for structure formation, the scenario of particles undergoing weak interactions has rapidly become the leading one~\cite{Hut:1977zn,Lee:1977ua,Vysotsky:1977pe}. These particles would have been at thermal equilibrium during the first instants of the Universe before seeing their density freeze out when the expansion rate exceeded their production/annihilation rate. Already in the 1980s, when few tests of the electroweak physics were available, the phenomenology of the Standard Model (SM) of particle physics was anticipated to be more complex than nowadays. One issue was notably due to the presence of a scalar field, the Higgs boson, the mass propagator of which leads to quadratic divergences in terms of the UV cut-off $\Lambda_\mathrm{UV}$. This cut-off can be identified with the threshold beyond which the complete mass spectrum of an underlying theory, more fundamental than the SM, contains one or more particles with a mass scale of order $\Lambda_\mathrm{UV}$. These particles regulate the effective low-energy theory by acting as cut-offs in the loop diagrams, but they can also de-stabilize the theory. At the order of one loop indeed, the Higgs mass receives quadratic corrections in $\Lambda_\mathrm{UV}$, dominated by the contribution in $m_\mathrm{t}^2/v^2$ of the heaviest fermion, the top quark, with $m_\mathrm{t}$ the mass of the top and $v$ the vacuum expectation value. Under these conditions, the corrections can rapidly explode with respect to the Higgs mass so that a fine tuning of the bare-mass parameter becomes necessary. Such a fine tuning goes against the prejudice that the mass inferred from a theory should be stable under small variations of the fundamental bare-mass parameter. The argument of naturalness was specifically introduced to determine the range of $\Lambda_\mathrm{UV}$ to avoid such fine tuning~\cite{tHooft:1979rat}. 

Formally, naturalness states that for any parameter $p$ that is small compared to the fundamental scale $\Lambda_\mathrm{UV}$, the limit $p\rightarrow0$ corresponds to an enhancement of the symmetry of the theory. This is because such an enhancement prevents from important radiative corrections. In this case indeed, at the fundamental high-energy scale at which the running of the Higgs quartic coupling $\lambda$ explodes (the Landau pole), the limit $m_\mathrm{H}\rightarrow0$ (with $m_\mathrm{H}$ the mass of the Higgs) does not correspond to any symmetry enhancement if the energy scale $\Lambda_\mathrm{UV}$ is much larger than $m_\mathrm{H}/\lambda \sim v$. In other words, the SM is natural only up to a scale not larger than the TeV. Hence, various mechanisms have been thought to add to the spectrum of elementary particles other ones, one of which would be stable with a mass $M_X$ around $100~$GeV and weak couplings. Such a weakly-interacting massive particle (WIMP) would have a self-annihilation cross section proportional to $G_\mathrm{F}^2M_X^2$ (with $G_\mathrm{F}$ the Fermi constant), which would roughly equate that needed to get the correct abundance of DM today via thermal production. This seemingly effortless matching, often referred to as the ``WIMP miracle'', explains that the WIMP paradigm has been a compelling one to explain DM and has largely defined the roadmap of experimental searches during the last few decades. 

However, the LHC has so far given no sign of the existence of new degrees of freedom, although energies up to the TeV have been explored and the favored threshold of $100~$GeV has been passed. In parallel, since 1985, a frantic hunt for a direct detection of WIMP DM has been going on in underground tunnels under the mountains (Fr\'ejus, Grand Sasso, or the old gold mine in Dakota) so as to be screened from cosmic rays by kilometers of rocks. The detection principle is common to all experiments: the WIMP is expected to interact coherently with a nucleus of ordinary matter, and the recoiling nucleus leaves a tiny energy deposit in the detector as ionization, scintillation light, and phonons depending on the specific detector technique. During these years, the upper limits have come down by almost ten orders of magnitude, and one can already conclude that these experiments have been very successful in not detecting DM. The race to discover WIMPs is led by noble liquids experiments, which have incrementally reached $\sim 100~$kg to $1~$ton of active mass, with xenon~\cite{PandaX-II:2016vec,LUX:2016ggv,XENON:2017vdw} or argon~\cite{DarkSide:2018kuk}. The next generation of experiments will reach several tons of mass~\cite{LZ:2015kxe,XENON:2017lvq,DarkSide-20k:2017zyg}. Soon a lower limit will be reached, where detection is expected, but this time due to the neutrinos from the Sun scattering off the detector nuclei (neutrino floor). Besides, numerous indirect-detection astrophysical searches have also been pursued based on the WIMP annihilation in SM particles. Such searches are subject to many uncertainties, in particular possible backgrounds of astrophysical origin or modeling of DM halo profiles. Overall, the various null results push the originally expected masses towards larger values and the couplings towards weaker ones. This gives increasingly strong constraints for the WIMPs to match the relic density. 

Although the exploration of the complete WIMP parameter space remains of great importance for the DM experimental program, the halo around the ``WIMP miracle'' started to deflate and a broader search program is actively pursued. The study presented in this contribution to ERCS2022 explores, using the Pierre Auger Observatory, the constraints that can be inferred on super-heavy dark matter (SHDM) as well as on the cosmological aspects of their production~\cite{PierreAuger:2022wzk,PierreAuger:2022ibr}. If they abandon the argument of naturalness, the reasons that motivate today SHDM keep the perspective of the SM as a low-energy effective theory from which the fundamental scale $\Lambda_\mathrm{UV}$ can be inferred from the analysis of the energy scale $\Lambda_\mathrm{I}$ at which the SM Higgs potential develops an instability at large field values. For the current values of the Higgs and top masses and the strong coupling constant, the range of $\Lambda_\mathrm{I}$ turns out to be high, namely $10^{10}$ to $10^{12}$\,GeV~\cite{Buttazzo:2013uya,Alekhin:2012py,Bednyakov:2015sca}. While the change of sign of $\lambda$ at that scale could trigger a vacuum instability due to the Higgs potential suddenly becoming unbounded from below, the running of $\lambda$ for energies above $\Lambda_\mathrm{I}$ turns out to be slow~\cite{Buttazzo:2013uya}. This peculiar behaviour leaves the possibility of extrapolating the SM to even higher energies than $\Lambda_\mathrm{I}$, up to $M_\textrm{Pl}$, with no need to introduce new physics to stabilize the SM. In this case, the mass spectrum of the dark sector could reflect the high energy scale of the new physics. 

%Actually, the masses of the Higgs boson and of the top quark place the SM vacuum in the metastable region, at the boundary of the stable one. Still, a quantitative estimation of its rate of quantum tunnelling into a lower energy state leads to a lifetime comfortably larger than the age of the universe~\cite{Buttazzo:2013uya,Andreassen:2017rzq}. This is without considering the cosmological context, in which the decay rate could have been amplified due to a high Hubble rate during inflation and to high temperature afterwards. Yet, it has been shown that the SM vacuum retains its astronomically-long lifetime for viable values of the Hubble rate $H_{\mathrm{inf}}$ and the non-minimal coupling $\xi$ between the Higgs field and the curvature of space-time~\cite{Markkanen:2018bfx}. 

Various mechanisms taking place at the end of the inflationary era in Big Bang cosmology are capable of producing SHDM particles. Inflation could be driven by the presence of a scalar field, the inflaton, which slowly rolled down its potential during the inflationary era before reaching its minimum. The inflaton field then started coherent oscillations  around its minimum potential and subsequently decayed into SM particles that reheated the universe (the reheating era) while thermalizing. We consider in this contribution the production of SHDM that could have occurred during this period of reheating, by annihilation of SM particles through the exchange of a graviton~\cite{Garny:2015sjg}, or by annihilation of inflaton particles through the same exchange of a graviton~\cite{Mambrini:2021zpp}. In this context, the only interaction between SM and dark sectors is  gravitational. The absence of DM-SM couplings is consistent with the large panoply of observational evidence for the existence of DM based on gravitational effects alone. Once SM and inflaton particles have populated the dark sector prior to the radiation-dominated era, the abundance of particles can evolve to match the relic abundance of DM inferred today for viable parameters governing the thermal history and geometry of the universe~\cite{Garny:2015sjg,Mambrini:2021zpp}. 

The absence of direct coupling between with SM particles (apart from gravitational) leaves only a few possible observational signatures. The large values of the Hubble expansion rate at the end of inflation $H_\textrm{inf}$ needed to match the relic abundance $\Omega_\text{CDM}h$ imply tensor modes in the cosmological microwave background anisotropies that could be observed in the future~\cite{Garny:2015sjg}. On the other hand, even if the absence of direct interactions guarantees the stability of the particles in the perturbative domain, SHDM protected from decay by a symmetry can eventually disintegrate due to non-perturbative effects in non-abelian gauge theories and produce ultra-high energy cosmic rays (UHECRs) such as (anti-)protons/neutrons, photons and (anti-)neutrinos. The aim of this study is to report on such signatures in the data from the Pierre Auger Observatory and to derive constraints on the various particle-physics and cosmological parameters governing the viability of the SHDM scenario. 

Instantons are a special class of field strengths in non-abelian gauge theories that satisfy the Yang-Mills equation of motion in Euclidean space~\cite{Belavin:1975fg,Coleman:1978ae,Vainshtein:1981wh}. The associated potentials tend to be pure gauge-field configurations that have the vacuum, or any gauge-rotated state thereof, at the boundary. The underlying gauge transformations cannot be continuously deformed to the identity. Therefore, the gauge vacua cannot be deformed into each other via a series of infinitesimal gauge rotations and fall into distinct classes, which are labeled by a topological quantum number known as the winding number (a.k.a.~Pontryagin index). This number characterizes the long-range structure of gauge field configurations, which is connected to their local properties associated with ultra-violet divergences during the renormalization step. This gives rise to anomaly relationships that reflect the breaking of classical symmetries and induce the non-conservation at the quantum level of certain currents that are not associated with gauge interactions~\cite{Adler:1969gk,Bell:1969ts}. In this way, exchanges of quantum numbers otherwise forbidden in the pertubative domain can take place as long as they are accompanied by changes in field configurations and thus in winding numbers. Instantons thus provide a signal for the occurrence of quantum tunneling between distinct classes of vacua~\cite{tHooft:1976rip}. 

Instanton-induced decay can therefore make observable a dark sector that would otherwise be totally hidden by the conservation of a quantum number~\cite{Kuzmin:1997jua}. Assuming quarks and leptons carry this quantum number and so contribute to anomaly relationships with contributions from the dark sector, they will be by-product decays together with the lightest hidden fermion. The lifetime of the decaying particle follows from Ref.~\cite{tHooft:1976rip},
\begin{equation}
\label{eqn:tauX}
\tau_X \simeq M_X^{-1}\exp{\left(4\pi/\alpha_X\right)},
\end{equation}
with $\alpha_X$ the reduced coupling constant of the hidden gauge interaction. In this expression, we retained only the exponential dependency in $\alpha_X^{-1}$, dropping the functional  determinants arising from the effect of quantum  fluctuations around the (classical) contribution of the instanton configurations as well as other functional determinants arising from the exact content of fields of the underlying theory. The constraints inferred using Eq.~\eqref{eqn:tauX} are indeed barely destabilized for a wide range of numerical factors given the exponential dependency in $\alpha_X^{-1}$. Eq.~\eqref{eqn:tauX} provides us with a relationship connecting the lifetime $\tau_X$, which is shown below to be constrained by the absence of UHE photons, to the coupling constant $\alpha_X$. 

In most SHDM models, the production of quark/anti-quark pairs is expected in the decay by-products, giving rise to large fluxes of UHE photons, $\dif N_\gamma/\dif E$. This is because each pair triggers a QCD cascade until the hadronization of the partons occurs and the unstable hadrons eventually decay. The flux of photons is calculated in terms of $x=2E/M_X$ as~\cite{Sarkar:2001se,Aloisio:2003xj}
\begin{equation}
    \frac{\dif N_\gamma(x)}{\dif x}=\frac{n(n-1)(n-2)\epsilon_\pi}{3}\int_x^1 \frac{\dif z}{z} \frac{x}{z}\left(1-\frac{x}{z}\right)^{n-3}\frac{D_h(z)}{z},
\end{equation}
with $n$ the (even) multiplicity of $q\bar q$ pairs, $\epsilon_\pi$ the ``efficiency'' of the hadronization process into pions, and $D_h(z)$ the function of fragmentation of a parton into a hadron obtained using the computational scheme described in Ref.~\cite{Aloisio:2003xj}. Due to their attenuation over intergalactic distances, only those photons emitted in the Milky Way can survive on their way to Earth. The emission rate per unit volume and unit energy $q_\gamma$ from any point labelled by its Galactic coordinates is shaped by the density of DM particles, $n_\text{DM}$,
\begin{equation}
    \label{eqn:q_gamma}
    q_\gamma(E,\mathbf{x}_\odot+s\nn) = \frac{1}{\tau_X}\frac{\dif N_\gamma}{\dif E}n_\text{DM}(\mathbf{x}_\odot+s\nn),
\end{equation}
where $\tau_X$ is the lifetime of the particle, $\mathbf{x}_\odot$ is the position of the Solar system in the Galaxy, $s$ is the distance from $\mathbf{x}_\odot$ to the emission point, and $\nn\equiv\nn(\ell,b)$ is a unit vector on the sphere pointing to the Galactic longitude $\ell$ and latitude $b$. Hereafter, the density is more conveniently expressed in terms of energy density $\rho_\text{DM}(\mathbf{x})=M_Xn_\text{DM}(\mathbf{x})$, normalized to $\rho(\mathbf{x}_\odot)=0.3$\,GeV\,cm$^{-3}$. There are uncertainties in the determination of this profile. We take as reference the traditional NFW profile~\cite{Navarro:1995iw}. 

The expected flux (per steradian) of UHE photons produced by the decay of super-heavy particles, $J_{\text{DM},\gamma}(E,\nn)$, is obtained by integrating the position-dependent emission rate $q_\gamma$ along the path of the photons in the direction $\nn$,
\begin{equation}
    \label{eqn:Jgal}
    J_{\mathrm{DM},\gamma}(E,\nn)=\frac{1}{4\pi}\int_0^\infty \dif s~q_\gamma(E,\mathbf{x}_{\odot}+s\nn),
\end{equation}
where the $4\pi$ normalization factor accounts for the isotropy of the decay processes. While the peak value of the flux is inversely proportional to the unknown $M_X$ and $\tau_X$ parameters, the energy and directional dependencies are determined by Eq.\eqref{eqn:q_gamma}. The exact content of quarks and leptons  depends on the specific underlying model. Yet, instanton-induced decays obey selection rules that involve necessarily large multiplicities. As a proxy, we consider a dozen of $q\overline{q}$ pairs. The flux pattern in the sky is more intense in a hot-spot region around the Galactic center; it provides therefore clear signatures. On the other hand, the non-observation of UHE photons reported in~\cite{PierreAuger:2022uwd,Savina:2021ufs,PierreAuger:2022aty} enables one to constrain the all-sky flux observed over the solid angle $\Delta\Omega$, $\langle J_{\mathrm{DM},\gamma}(E,\nn)\rangle=\int_{\Delta\Omega}\dif\nn ~J_{\mathrm{DM},\gamma}(E,\nn)/\Delta\Omega$, and thus to constrain the unknown $M_X$ and $\tau_X$ parameters.

\begin{figure}[t]
\centering
\includegraphics[width=0.83\columnwidth]{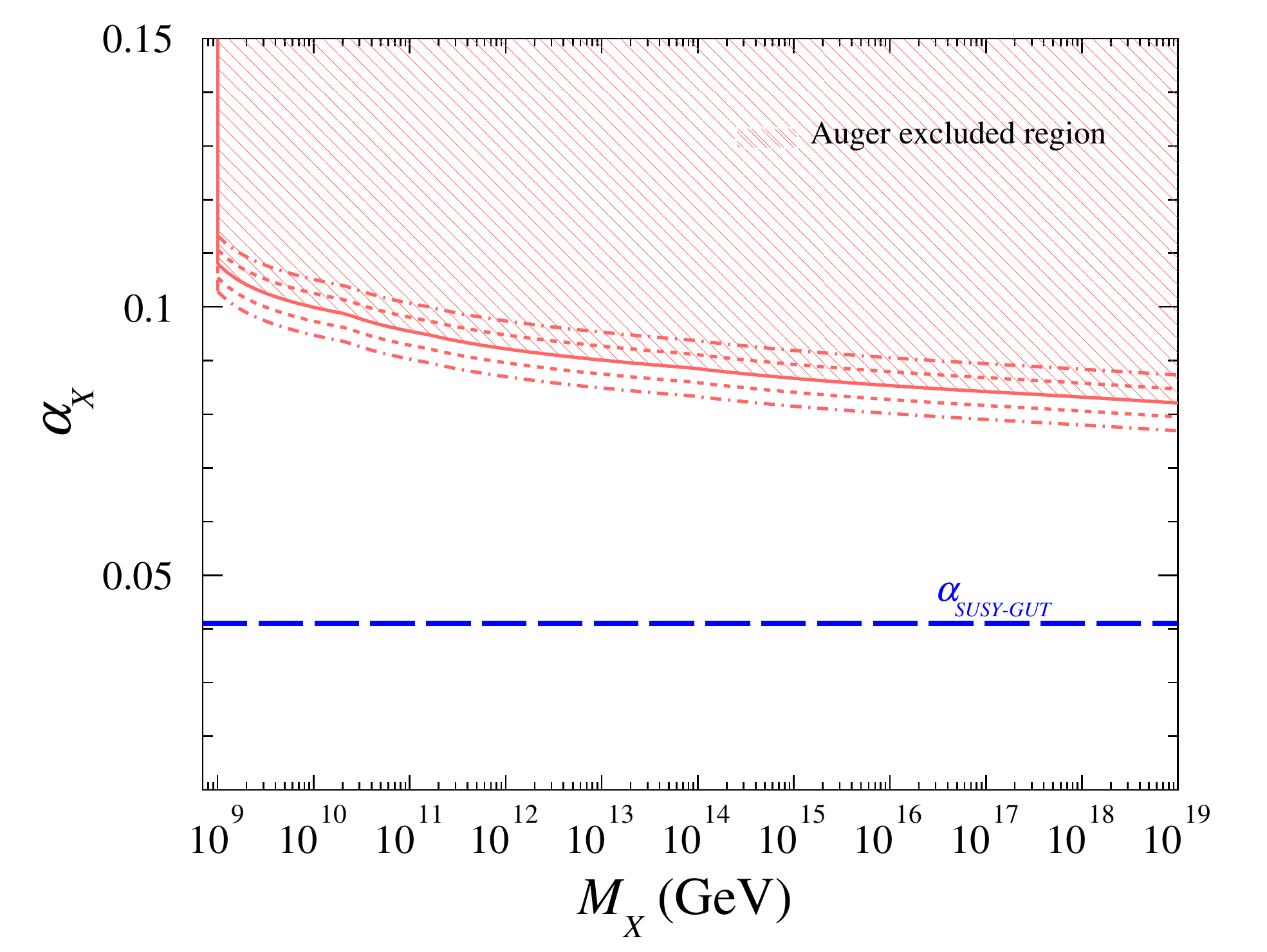}
\caption{Upper limits at 95\% C.L.\ on the  coupling constant $\alpha_X$ of a hidden gauge interaction as a function of the mass $M_X$ of a dark matter particle decaying into a dozen of $q \bar q$ pairs. For reference, the unification of the three SM gauge couplings is shown as the blue dashed line in the framework of supersymmetric GUT~\cite{10.1093/ptep/ptaa104}.}
\label{fig:alphaX-mass}
\end{figure}

Assuming that the relic abundance of DM is saturated by super-heavy particles, constraints can be inferred in the plane $(\tau_X,M_X)$ by requiring the all-sky flux integrated above some energy threshold $E$ to be less than the limits, $\int_E^\infty\dif E'\langle J_{\text{DM},\gamma}(E',\nn)\rangle \leq J^{95\%}_\gamma({\geq}E)$. For a specific upper limit at one energy threshold, a scan of the value of the mass $M_X$ is carried out so as to infer a lower limit of the $\tau_X$ parameter, which is subsequently transformed into an upper limit on $\alpha_X$ by means of Eq.~\eqref{eqn:tauX}. This defines a curve. By repeating the procedure for each upper limit on $J^{95\%}_\gamma({\geq}E)$, a set of curves is obtained, reflecting the sensitivity of a specific energy threshold to some range of mass. The union of the excluded regions finally provides the constraints in the plane $(\alpha_X,M_X)$. In this manner the shaded red area is obtained in Fig.~\ref{fig:alphaX-mass}. As already noted, additional factors coming from the functional integral over the quantum fluctuations to describe the vacuum transition amplitude could be at play in Eq.~\eqref{eqn:tauX}~\cite{tHooft:1976snw}. Explicit constructions of the dark sector are required to calculate these factors. Such constructions are well beyond the scope of this study. Although the limits presented in Fig.~\ref{fig:alphaX-mass} are hardly destabilized due to the exponential dependence in $\alpha_X^{-1}$, we note that a shift of $\pm 0.0013k$ for factors $10^{\pm k}$ and limit ourselves to showing in dotted and dashed lines the bounds that would be obtained for $k=2$ and $k=4$, respectively. These factors are by far the dominant systematic uncertainties.

In summary, the minimal setup to produce DM is from gravitational effects alone, consistent with the concordance model of cosmology. The only unambiguous signature to capture the existence of such DM is through the detection of UHE photons produced by the instanton-induced decay. The non-observation of such fluxes allowed us to probe in a unique way to date the instanton strength through the dark-sector gauge coupling. There are some connection of these results to cosmological scenarios. Details can be found in Ref~\cite{PierreAuger:2022ibr}. 

%\begin{thebibliography}{99}
\bibliographystyle{JHEP}
\bibliography{bibliography}

%\end{thebibliography}

\end{document}